\documentstyle[graphics,twocolumn]{mn}
\newcommand{\mincir}{\raise
-2.truept\hbox{\rlap{\hbox{$\sim$}}\raise5.truept 
\hbox{$<$}\ }}
\newcommand{\magcir}{\raise
-2.truept\hbox{\rlap{\hbox{$\sim$}}\raise5.truept
\hbox{$>$}\ }}
\newcommand{\minmag}{\raise-2.truept\hbox{\rlap{\hbox{$<$}}\raise
6.truept\hbox
{$>$}\ }}

\newcommand{\lya}{Lyman-$\alpha$~}

\newcommand{\vm}{{\rm s/km}}

\newcommand{\be}{\begin{equation}}
\newcommand{\ee}{\end{equation}}
\newcommand{\ba}{\begin{eqnarray}}
\newcommand{\ea}{\end{eqnarray}}
\newcommand{\brr}{\begin{array}}
 
\newcommand{\err}{\end{array}}
\newcommand{\bc}{\begin{center}}
\newcommand{\ec}{\end{center}}

\newcommand{\mpch} {\rm $h^{-1}$ Mpc\,\,}

\DeclareMathAlphabet{\mathsc}{OT1}{cmr}{m}{sc}
\def\testbx{bx}%
\DeclareRobustCommand{\ion}[2]{%
\relax\ifmmode
\ifx\testbx\f@series
{\mathbf{#1\,\mathsc{#2}}}\else
{\mathrm{#1\,\mathsc{#2}}}\fi
\else\textup{#1\,{\mdseries\textsc{#2}}}%
\fi}
\title[Inferring the linear  matter  power spectrum from 
the  \lya forest]{Inferring
the dark matter  power spectrum from the  \lya forest in
high-resolution QSO absorption spectra}

\author[M. Viel, M.G. Haehnelt \& V. Springel] {Matteo Viel$^{1}$,
Martin G. Haehnelt$^{1}$ \& Volker Springel$^{2}$ \\ $^1$ Institute of
Astronomy, Madingley Road, Cambridge CB3 0HA\\ $^2$
Max-Planck-Institut f\"ur Astrophysik, Karl-Schwarzschild-Str. 1,
Garching bei M\"unchen, Germany \\ \\}

\begin{document}

\maketitle
\begin{abstract}

We use the LUQAS sample (Kim et al. 2004), a set of 27 high-resolution
and high signal-to-noise QSO absorption spectra at a median redshift
of $z=2.25$, and the data from Croft et al. (2002) at a median
redshift of $z=2.72$, together with a large suite of high-resolution
large box-size hydro-dynamical simulations, to estimate the linear
dark matter power spectrum on scales $0.003\,{\rm s/km} < k <
0.03\,{\rm s/km}$.  Our re-analysis of the Croft et al.~data agrees
well with their results if we assume the same mean optical depth and
gas temperature-density relation.  The inferred linear dark matter
power spectrum at $z=2.72$ also agrees with that inferred from LUQAS
at lower redshift if we assume that the increase of the amplitude is
due to gravitational growth between these redshifts.  We further argue
that the smaller mean optical depth measured from high-resolution
spectra is more accurate  than the larger value obtained from
low-resolution spectra by Press et al.~(1993) which Croft et
al.~used. For the smaller optical depth we obtain a $\approx 20$\%
higher value for the rms fluctuation amplitude of the matter
density. By combining the amplitude of the matter power spectrum
inferred from the Ly$\alpha$ forest with the amplitude on large scales
inferred from measurements of the CMB we obtain constraints on the
primordial spectral index $n$ and the normalisation $\sigma_8$.
For values of the mean optical depth favoured by high-resolution
spectra, the inferred linear power spectrum is consistent with a
$\Lambda$CDM model with a scale-free ($n=1$) primordial power
spectrum.
\end{abstract}


\section{Introduction}

The prominent absorption features blue-ward of the \lya emission in the
spectra of high-redshift quasars (QSOs) are now generally believed to
arise from smooth density fluctuations of a photoionised warm
intergalactic medium (see Rauch 1998 and Weinberg et al. 1999 for
reviews).  This has opened up the possibility to probe the density
fluctuation of matter with the flux power spectrum of QSO absorption
lines (Croft et al. 1998, Hui 1999, Croft et al. 1999b, McDonald et al. 2000
[M00], Hui et al. 2001, Croft et al. 2002 [C02], McDonald 2003, Viel et al. 2003). 

The flux power spectrum is mainly sensitive to the slope and amplitude
of the linear dark matter power spectrum for wave-numbers in the range
$0.002\, \vm <k< 0.05\, \vm$.  Croft et al. (1999) inferred an
amplitude and slope which was consistent with a COBE normalised
$\Lambda$CDM model with a primordial scale invariant fluctuation
spectrum (Phillips et al. 2001).  M00 and C02, using a larger sample
of better quality data, found a somewhat shallower slope and smaller
fluctuation amplitude.  The WMAP team used the later data in combination
with their CMB data to claim that there is evidence for a tilted primordial
CMB-normalised fluctuation spectrum ($n<1$) and/or a running spectral
index (Bennet et al. 2003; Spergel et al. 2003; Verde et al. 2003).  A
number of authors have argued that the errors in the inferred dark
matter (DM) power spectrum have been underestimated (Zaldarriaga,
Scoccimarro \& Hui, 2003; Zaldarriaga, Hui \& Tegmark, 2001; Gnedin \&
Hamilton 2002; Seljak, McDonald \& Makarov 2003).  We here use a suite
of high-resolution hydro-dynamical simulations and the flux power
spectrum obtained from LUQAS (Large Sample of UVES QSO Absorption
Spectra), together with the published flux power spectrum of C02 to
further investigate these issues.

The plan of the paper is as follows. In Section \ref{data} we describe
the two data sets and give an overview of the determinations of the
effective optical depth found in the literature.  The flux power
spectra obtained from the hydro-dynamical simulations are discussed in
Section \ref{hydro}.  Section \ref{inferring} describes the method and
uncertainties of inferring the linear matter power spectrum.  In
Section \ref{linear} we present our results, compare with previous
results and discuss implications for $\sigma_8$ and $n$. Section
\ref{conclu} contains a summary and our conclusions.

\section{The observed flux power spectrum and effective optical depth}
\label{data}
We will use estimates of the flux power spectrum from two different
data sets: the LUQAS sample of VLT spectra compiled by K04 and the
sample of Keck spectra compiled by C02. Below, we will describe both
samples in more detail.

\subsection{The LUQAS sample} 
\label{luqas}

LUQAS (Large Sample of UVES Quasar Absorption Spectra) compiled by
K04, consists of 27 high-resolution spectra taken with the VLT UVES
spectrograph.  The median redshift of the \lya forest probed by the
spectra is $\left<z\right>=2.25$ and the total redshift path covered
is $\Delta z = 13.75$.  The S/N varies as a function of wavelength,
but in the forest region it is usually larger than 50. The flux power
spectrum of the LUQAS sample has been analysed in K04, while a study
of the flux bispectrum can be found in Viel et al. (2004b).  

The LUQAS sample probes somewhat smaller redshifts than the C02
sample, but, as discussed by K04, in the redshift range where the two
samples overlap, the estimated flux power spectra agree 
well in the range of wavenumbers not affected by differences in resolution and S/N.  
Here, we will use a subset of the LUQAS sample for which we
have selected all QSO spectra regions in the redshift range $2<z<2.3$,
to maximise the contrast in redshift to the C02 sample and to further
investigate the redshift evolution.  16 QSOs contribute to this
subsample (Figure~1 in K04) and the median redshift is
$\left<z\right>=2.125$.  In Table~\ref{tab1}, we give the 3D flux
power spectrum $\Delta^2_F(k)=P_{F}^{\rm 3D}(k)\,k^3/(2\pi^2)$ for this
sample, using the flux estimator $F/\left<F\right>-1$ (denoted $F2$ in
K04), where $F$ is the flux of the continuum-fitted spectra.  The 3D
flux power spectrum is obtained from the 1D flux power spectrum by
differentiation in the usual way, viz.\be P^{\rm 3D}_{F} =
-\frac{2\,\pi}{k}\frac{{\rm d}P^{\rm 1D}_{F}}{{\rm d}k}\; .  \ee

Note that peculiar velocities and thermal broadening make the flux
field anisotropic and that $P_{F}^{\rm 3D}$ is thus not the true 3D
power spectrum of the flux. The flux power spectrum has been
calculated for the same wavenumbers as in C02.

\begin{figure}
\center\resizebox{0.5\textwidth}{!}{\includegraphics{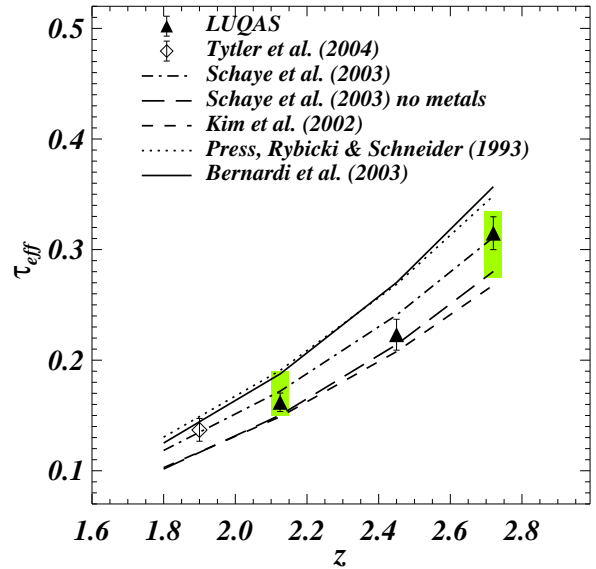}}
\caption{Triangles show the effective optical depth $\tau_{\rm
eff}=-\ln\left<F\right>$ for the LUQAS sample (also given in
Table 2).  The dashed curve is the result of Kim et
al.~(2002) while the long-dashed and dot-dashed curves show the result
of Schaye et al. (2003), with and without the removal of pixels
contaminated by metals, respectively.
The dotted curve is the effective optical depth obtained by Press,
Rybicki \& Schneider (1993). The solid curve shows the results from
Bernardi et al. (2003).  The empty diamond is the result of Tytler et
al.~(2004). The vertical extent of the two shaded regions indicates
the values used in our analysis. }
\label{fig1}
\end{figure}

\subsection{The Croft et al. sample}

The Croft et al. (2002) sample (C02) consists of 30 Keck HIRES spectra
and 23 Keck LRIS spectra.  We will here use the flux power spectrum
which C02 derived for what they call their `fiducial' sample. This
sample has a median redshift $\left<z\right>=2.72$, spans the redshift
range $2.3<z<3.2$, and has a total redshift path of $\Delta z =
25$. For more details see C02.

\begin{table}
\caption{The 1D  and 3D  power spectrum of the flux  $F=\exp(-\tau)/\left<\exp(-\tau)\right>-1$
of the LUQAS sample at $z=2.125$.}
\label{tab1}
\begin{tabular}{lcc}
\hline
\noalign{\smallskip}
$k$ (s/km) & $P^{\rm 1D}_{F}(k)$ (s/km) &  $\Delta^{2}_{F}(k)$ \\
\noalign{\smallskip}
\hline
\noalign{\smallskip}
 0.00199 & 18.4942 $\pm$  2.9106 &0.0002 $\pm$  0.0031\\
 0.00259 & 16.6206 $\pm$  2.7119 &0.0003 $\pm$  0.0040\\
 0.00336 & 21.0862 $\pm$  3.3361 &0.0004 $\pm$  0.0052\\
 0.00436 & 16.2223 $\pm$  1.7413 &0.0106 $\pm$  0.0039\\
 0.00567 & 13.9062 $\pm$  1.2669 &0.0153 $\pm$  0.0075\\
 0.00736 & 12.9220 $\pm$  2.3890 &0.0194 $\pm$  0.0077\\
 0.00956 &  9.6921 $\pm$  0.9864 &0.0156 $\pm$  0.0065\\
 0.01242 &  8.9783 $\pm$  0.7195 &0.0253 $\pm$  0.0065\\
 0.01614 &  7.2158 $\pm$  0.5574 &0.0421 $\pm$  0.0037\\
 0.02097 &  4.4987 $\pm$  0.2800 &0.0499 $\pm$  0.0054\\
 0.02724 &  3.3617 $\pm$  0.2313 &0.0468 $\pm$  0.0055\\
 0.03538 &  2.1198 $\pm$  0.1135 &0.0498 $\pm$  0.0054\\
 0.04597 &  1.1490 $\pm$  0.0677 &0.0424 $\pm$  0.0032\\
 0.05971 &  0.6303 $\pm$  0.0480 &0.0345 $\pm$  0.0030\\
 0.07757 &  0.2833 $\pm$  0.0169 &0.0256 $\pm$  0.0022\\
\noalign{\smallskip}
\hline
\end{tabular}
\end{table}

\begin{table}
\caption{Mean flux and effective optical depth of the LUQAS sample.}
\label{tab2}
\begin{tabular}{cccc}
\hline
\noalign{\smallskip}
redshift range&$\left<z\right>$&$\left<F\right>$ & $\tau_{\rm eff}$ \\
\noalign{\smallskip}
\hline
\noalign{\smallskip}
$ 2.0<z<2.3$&2.125  & $ 0.849 \pm 0.008 $ &  $0.163 \pm {0.009}$ \\ 
$ 2.3<z<2.6$& 2.44  & $ 0.800 \pm 0.008 $ &  $0.223 \pm {0.014}$ \\
$ 2.55<z<3.0$&2.72 & $0.730 \pm 0.011 $ &   $0.315 \pm {0.015}$\\ 
\noalign{\smallskip}
\hline
\end{tabular}
\end{table}


\subsection{Statistical and systematic errors in the observed flux power spectrum}  
\label{errors}

Unless stated otherwise, we estimate our statistical errors with a
jack-knife estimator.  The main systematic errors affecting estimates
of the flux power spectrum due to absorption by the \lya forest are
continuum fitting and the presence of metal lines and damped
Lyman-$\alpha$ systems. These effects have been investigated by
e.g. C02 and K04. The main conclusions of K04 are as follows: {\it i)}
continuum fitting uncertainties for high-resolution Echelle spectra
strongly affect the flux power spectrum at scales $k<0.003\,{\rm
s/km}$; {\it ii)} the contribution of metal lines is less than 10\% at
scales $k<0.01\,{\rm s/km}$ but rises significantly (up to 50\%) at
smaller scales; {\it iii)} damped Lyman-$\alpha$ systems appear to
only mildly affect the estimate of the flux power spectrum.  In order
to minimise uncertainties due to continuum fitting and metal lines,
and to avoid dealing with the problematic thermal cut-off at small
scales, we will only use the range of wavenumbers $0.003 < k  \,{\rm
(s/km)} <0.03$ for our analysis.

\subsection{The observed effective optical depth} 
\label{effopt}

As pointed out by Croft et al. (1998), C02 and Seljak, McDonald \&
Makarov (2003) and discussed in detail in Sections \ref{inferring} and
\ref{statistical}, the assumed effective optical depth, $\tau_{\rm
eff} = - \ln{\left<F\right>}\,$, has a large influence on the
amplitude of the inferred matter power spectrum. The observed values
of $\tau_{\rm eff}$ have large statistical and probably also not yet
fully understood systematic errors.  The main uncertainty in
determining the effective optical depth comes from the continuum
fitting procedure and the Poisson noise due to the large variations
from line-of-sight to line-of-sight (Zuo \& Bond 1994, Viel
et al. 2004a, Tytler et al. 2004).  

In Figure \ref{fig1}, we show results for the redshift evolution of
$\tau_{\rm eff}$ from a number of observational studies in the
literature. The dotted curve is from a large sample of low-resolution
spectra compiled by Press, Rybicki \& Schneider (1993, PRS), while the
solid line is from a sample of low resolution spectra drawn from SDSS
by Bernardi et al. (2003).  The open diamond shows the result from
Tytler et al. (2004).  The short-dashed curve is the result of Kim et
al. (2002) using a sample of high-resolution UVES spectra, while the
long-dashed and dash-dotted curves show the results of Schaye et
al.~(2003) from a combined sample of high-resolution UVES and Keck
spectra with and without the removal of pixels contaminated by
associated metal absorption.  The filled triangles show the results
from the LUQAS sample for three different redshift ranges (also given
in Table \ref{tab2}).  Damped and sub-damped \lya systems have been
removed for the estimate from the LUQAS sample and errors have been
calculated with a jack-knife estimate using 40 subsamples. Note that
the sample used by Schaye et al.~(2003) has 14 QSO spectra in common with the
LUQAS sample.

The results of PRS and Bernardi et al.~(2003), which both use
low-resolution spectra with comparably low S/N, give results which are
about 15-20\% higher than those of studies using high-resolution
spectra (see also Schaye et al. 2003 for a discussion). As argued by
Seljak et al. (2003) and Tytler et al. (2004) this is most likely due
to systematic errors in the continuum fitting procedure of the
low-resolution spectra.  We agree with this assessment.  At redshifts
$\le 2.7$ the expected continuum fitting errors in high-resolution
spectra is of order 2\% (Bob Carswell and Tae-Sun Kim, private
communication). 
For spectra at these redshifts, the errors are thus almost certainly
dominated by the somewhat uncertain contribution from metal lines and
statistical errors.  Based on Figure \ref{fig1} we will adopt the
following values for the optical depth and its error in our further
analysis, $\tau_{\rm eff} (z=2.125) =0.17\pm 0.02$ and $\tau_{\rm eff}
(z=2.72) = 0.305 \pm 0.030$.

\section{The flux power spectrum of simulated absorption spectra} 
\label{hydro}
\subsection{Numerical code and parameters} 
\begin{figure*}
\center\resizebox{1.0\textwidth}{!}{\includegraphics{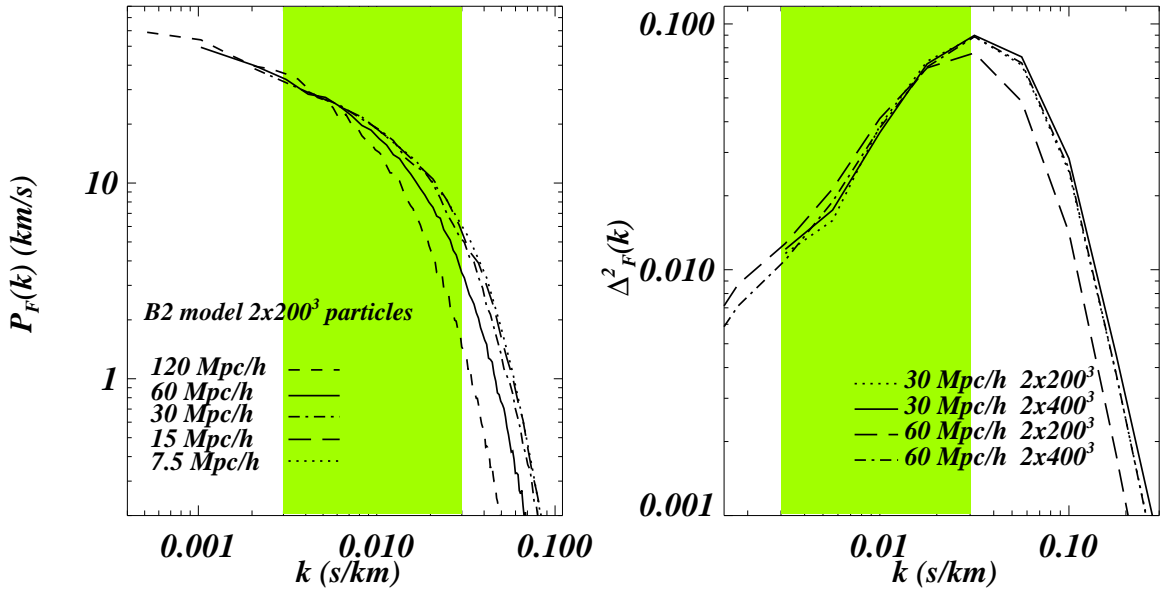}}
\caption{{\it Left panel:} 1D flux power spectra of different
simulations of model B2 at $z=2.75$, run with different box-sizes. All
simulations are with 200$^3$ dark matter particles and 200$^3$ gas
particles.  Note that simulations have not been scaled to the same
effective optical depth. The shaded region indicates the range of
wavenumbers used to infer the linear dark matter power spectrum.  {\it
Right panel:} 3D flux power spectra for different simulations of model
B2, for a range of box-sizes and resolutions, as indicated in the
plot.}
\label{fig2}
\end{figure*}

\begin{table}
\caption{Grid of cosmological simulation parameters.}
\label{tab4}
\begin{tabular}{lcccc}
\hline
\noalign{\smallskip}
$\;\;$&  $\sigma_8=0.7$ & $\sigma_8=0.85$ & $\sigma_8=1.0$ \\
\noalign{\smallskip}
\hline
\noalign{\smallskip}
$ n=0.95 $ & B1 & B2 & B3 \\
$ n=1.0 $  & C1 & C2 & C3 \\
\noalign{\smallskip}
\hline
\end{tabular}
\end{table}

We have run a suite of simulations with varying cosmological
parameters, particle numbers, resolution, boxsize and thermal
histories using a new version of the parallel TreeSPH code {\small
GADGET} (Springel, Yoshida \& White, 2001).  {\small GADGET-2} was
used in its TreePM mode which speeds up the calculation of long-range
gravitational forces considerably. The simulations were performed with
periodic boundary conditions with an equal number of dark matter and
gas particles and used the conservative `entropy-formulation' of SPH
proposed by Springel \& Hernquist (2002). Radiative cooling and
heating processes were followed using an implementation similar to
that of Katz et al.~(1996) for a primordial mix of hydrogen and
helium. We have assumed a mean UV background produced by quasars as
given by Haardt \& Madau (1996), which leads to reionisation of the
Universe at $z\simeq 6$, but have also run simulations where we
artificially increased the heating rates to mimic different thermal
histories.  Most simulations were run with heating rates increased by
a factor of 3.3 in order to achieve temperatures which are close to
observed temperatures (Schaye et al. 2000; Ricotti et al. 2000;
Choudhury, Srianand \& Padmanabhan 2001).

In order to maximise the speed of the simulations we have employed a
simplified star-formation criterion in the majority of our
runs. All gas at densities larger than 1000 times the
mean density was turned into collisionless stars. The absorption
systems producing the \lya forest have small overdensity so this
criterion has little effect on flux statistics, while speeding up the
calculation by a factor of $\sim 6$, because the small dynamical times
that would otherwise arise in the highly overdense gas need not to be
followed any more.  In a pixel-to-pixel comparison with a simulation
which adopted the full multi-phase star formation model of Springel \&
Hernquist (2003) we explicitly checked for any differences introduced
by this approximation.  We found that the differences in the flux
probability distribution function were smaller than 2\%, while the
differences in the flux-power spectrum were smaller than 0.2 \%.  We
have also turned off all feedback options of {\small GADGET-2} in our
simulations.  The effect of feedback by galactic winds on the
statistics of the flux distribution is uncertain but is believed to be
small (e.g. Theuns et al. 2002, Bruscoli et al. 2003, Croft et
al. 2003, Kollmeier et al. 2003, Desjacques et al. 2004).

The simulations were all started at $z=99$ and we have stored 19
redshift outputs for each run, mainly in the redshift range
$1.5<z<3.5$.  The initial gas temperature was $T = 227$ K, and $40\pm
2$ SPH neighbours were used to compute physical quantities. The
gravitational softening was set to 2.5 $h^{-1}$ kpc in comoving units
for all particles.

We have run a suite of simulations with cosmological parameters close
to the values obtained by the WMAP team in their analysis of WMAP and
other data (Spergel et al. 2003), as shown in Table~\ref{tab4}.  Our
fiducial model is a `concordance' $\Lambda$CDM model with
$\Omega_{0{\rm m}}= 0.26$, $\Omega_{0\Lambda} = 0.74$, $\Omega_{0{\rm b}} = 0.0463$ and
$H_0=72\,{\rm km\,s^{-1}Mpc^{-1}}$ (B2 in Table \ref{tab4}).  The CDM
transfer functions of all models have been taken from Eisenstein \& Hu
(1999).

The simulations were run on COSMOS, a shared-memory Altix 3700 with
128 900 MHz Itanium processors hosted at the Department of Applied
Mathematics and Theoretical Physics (Cambridge), and a 64-node Beowulf
cluster with 64 1.2 GHz Sun processors hosted at the Institute of
Astronomy (Cambridge).  The majority of our simulations have been run
with $2 \times 400^3$ particles in a 60 \mpch box and they took about
280 hrs on 32 processors to reach $z=2$.

\subsection{The effect of resolution and box size on the flux power
spectrum} 

As discussed in K04 and C02 and in Section~\ref{errors}, the range of
wavenumbers $0.003 <k /{\rm (s/km)}<0.03$ is least affected by
systematic uncertainties. We will thus limit our attempts to infer the
linear dark matter power spectrum to this range, which is indicated by
the shaded region in Figure~\ref{fig2} and following figures.  In the
left panel of Figure \ref{fig2}, we plot the 1D flux power spectrum
for a suite of exploratory runs with fixed particle number $2\times
200^3$ for five different box sizes/resolutions.  All simulations are
for our fiducial cosmology B2 and have been run with the same phases.  
There is generally good agreement in
the overlap regions of the flux power spectrum.  At large wavenumbers,
the flux power spectrum for simulations with box size $\ge
60\,h^{-1}{\rm Mpc}$, where the gas particle mass is larger than $5
\times 10^7 \,h^{-1}{\rm M}_{\odot}$, is however clearly affected by
insufficient resolution.

The right panel of Figure \ref{fig2} shows the effect of increasing
the mass resolution by a factor of eight for the $60\,h^{-1}{\rm Mpc}$
and $30\,h^{-1}{\rm Mpc}$ boxes, corresponding to an increase of the
particle number to $2\times 400^3$. Plotted is the 3D power spectrum
which we will use to infer the linear dark matter power spectrum using
the `effective bias' method developed by C02.  There is good agreement
between the flux power spectra for $k > 0.008$ s/km, except at small
scales for the 60 \mpch box with only $2\times 200^3$ particles.  A
box size of 60 \mpch with $2\times 400^3$ particles appears to be a
suitable compromise for estimates of the DM power spectrum. It has
converged on small scales and will only be moderately affected by
cosmic variance on the largest scales (for more discussion see
Section~\ref{statistical} on errors).  For this choice, the mass per
dark matter particle is $2 \times 10^8\,h^{-1}{\rm M}_{\odot}$ and the
mass per gas particle is $4.3 \times 10^7\,h^{-1} M_{\odot}$. Note
that this is a significant improvement compared to the simulations of
C02 who used 10 dark matter simulations with a box size of 27.77 \mpch
and $160^3$ particles.

\begin{figure*}
\center\resizebox{1.0\textwidth}{!}{\includegraphics{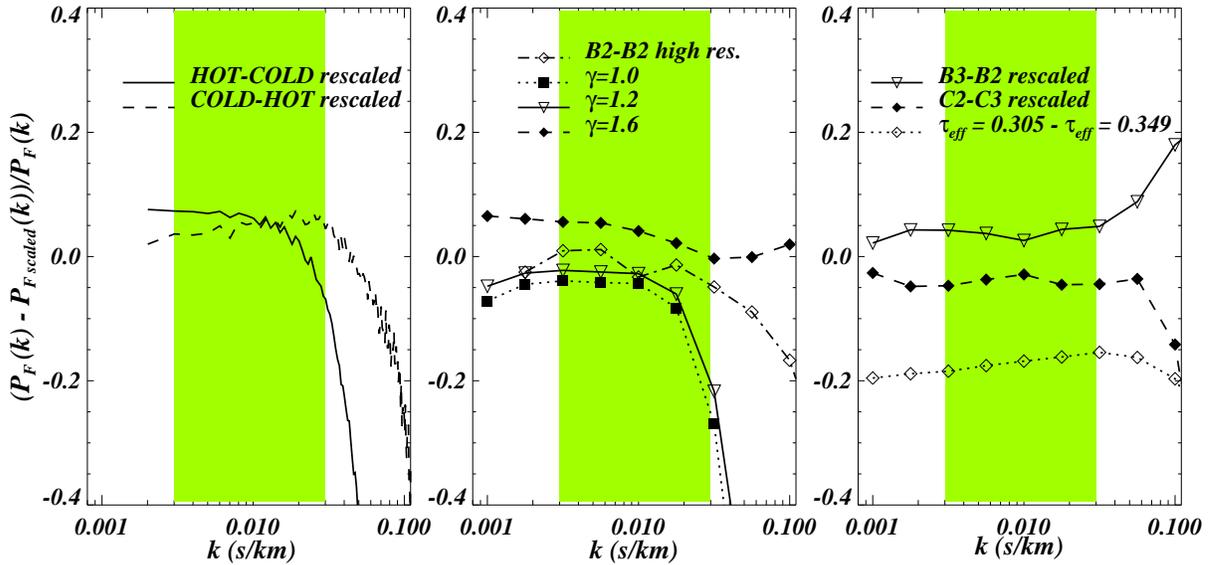}}
\caption{Effect of rescaling simulations (60 \mpch box $2\times 400^3$ particles),  to reproduce different
temperature-density relations and different values of $\sigma_8$.
{\it Left:} Difference between the 1D flux power spectrum of the `hot'
simulation and the rescaled `cold' simulation (solid curve) and
{\it vice versa} (dashed curve).  {\it Middle:} Difference
between the 1D flux power spectrum of the simulation of the B2 model
and the flux power spectra for three rescaled models with different
values of the exponent $\gamma$ of the temperature-density relation.
The difference between the 1D\, flux power spectra of two simulations of
model B2 with two different resolutions is also shown (30 \mpch box, $2\times 400^3$ particles).  {\it Right:} The
solid curve shows the difference between the 1D flux power spectra of
a simulation of model B2 ($\sigma_{8}=0.85, z=2.72$) and an output of
a simulation of model B3 ($\sigma_{8}=1$) at higher redshift which has
the same rms density fluctuation (rescaled to the effective optical
depth at $z=2.72$).  The dashed curve shows the difference between the
1D flux power spectra of a simulation of model C3 ($\sigma_{8}=1.0,
z=2.125$) and an output of a simulation of model C2
($\sigma_{8}=0.85$) at lower redshift which has the same rms density
fluctuation (rescaled to the effective optical depth at $z=2.125$).
The dotted curve shows the difference between the 1D flux power
spectra of a simulation of model B2 for two different values of
$\tau_{\rm eff}$. The shaded regions indicate the range of wavenumbers
used in our analysis.}
\label{fig3}
\end{figure*}

\subsection{The effect of temperature on the flux power spectrum 
   and the rescaling of the temperature-density relation} 

It is well established by analytical arguments and numerical
simulations that the gas responsible for the \lya forest is in
photoionisation equilibrium and exhibits a rather tight relation
between density and temperature (e.g. Hui \& Gnedin 1997). This
relation is a quasi-equilibrium established by the balance between
photo-heating and adiabatic cooling. For moderate over-density the
relation can be approximated by a power-law of the form,
\be
T = T_{0} \left ( \frac{\rho}{\left<\rho\right>} \right )^{\gamma-1}. 
\ee
The slope $\gamma$ and normalisation $T_{0}$ of this relation depend
on the reionisation history of the Universe.  The effect
on the flux power spectrum is twofold. Changing the temperature
changes the width of the absorption features and it thus changes the
mean flux decrement for a given distribution of neutral hydrogen.  

To see how the slope affects the flux power spectrum, it is helpful to
consider the fluctuating Gunn-Peterson approximation. The relation
between the optical depth and matter density in redshift space can be
written as
\begin{equation} 
\tau =  A  \left ( \frac{\rho}{\left<\rho\right>} \right ) ^{\beta},
\end{equation} 
where $\beta = 2.7-0.7\,\gamma$ depends on the temperature-density
relation of the gas due to the temperature dependence of the
recombination coefficient (see Weinberg et al. 1999 for a review). The
factor $A$ depends on redshift, baryon density, temperature at the
mean density, Hubble constant and photoionisation rate.  Increasing
the slope of the temperature-density relation therefore flattens the
slope of the relation between optical depth and matter density.  At
fixed mean optical depth, this results in a decrease of the amplitude
of the flux power spectrum.

With a running time of two weeks for each of our simulations we could
not afford to run an extensive parameter study of simulations with
different thermal histories and temperature-density relations.  We
have thus used the rescaling method developed by Theuns et al. (1998)
to impose a variety of temperature-density relations. This will not
account for the effects that the corresponding change in the gas
pressure would have had on the gas distribution. It nevertheless
mimics the major effects of different density temperature relations on
the flux power spectrum reasonably well. To verify this explicitly, we
have compared two simulations  run with a factor $\sim$ 3.3 different 
photo-heating rates. The  temperatures  were
different  by a  factor $\sim 2.5$.  The best fitting parameters for the
temperature-density relation of the `hot' simulation are
$T_0=10^{4.15}$ K and $\gamma=1.6$.

In the left panel of Figure \ref{fig4}, we compare the flux power
spectra of the `hot' and `cold' simulations with flux power spectra
for which the temperatures of the hot/cold simulations have been
rescaled such that they have the same temperature-density relation as
the cold/hot simulations.  The differences between the 1D flux power
spectra of rescaled and simulated models are smaller than 10\% at
small $k$, but start to diverge strongly at $k>0.02\,{\rm s/km}$. The
differences mostly depend only weakly on $k$ and the differences
between the 3D spectra will generally be smaller.  The middle panel
shows the effect of rescaling to temperature-density relations with
different $\gamma$.  We will explore the effect of changing the
temperature-density relation on the inferred dark matter power
spectrum in more detail in Section~\ref{inferring}.

\subsection{The amplitude of the matter power spectrum and rescaling of
the redshift}

\label{grid}
Our grid of simulations is somewhat sparse in the fluctuation
amplitude of the matter power spectrum. However, at the redshifts
considered here, redshift and fluctuation amplitude are largely
degenerate and a suitably rescaled simulation output from a different
redshift can mimic a simulation with different fluctuation
amplitude. In the right panel of Figure \ref{fig4} we test how well
this works by comparing the flux power spectrum of simulations with
different $\sigma_{8}$ at redshifts where the rms density fluctuation
amplitude of the simulation is the same.  The differences at large
scales ($k<0.03$ s/km) are typically of the order of 5\%.

\section{Estimating the matter power spectrum}
\label{inferring}
\subsection{The Method}
\label{method}

The method we use to infer the linear dark matter power spectrum has
been proposed by C02. It uses numerical simulations to calibrate the
relation between flux power spectrum and matter power spectrum.  It
then assumes that the flux power spectrum $P_F(k)$ at a given
wavenumber $k$ depends linearly on the linear real space matter power
spectrum $P(k)$ at the same wavenumber and that both can be related by
a simple bias function $b(k)$,
\be
P_F(k)=b^2(k) \,P_{\rm mat}(k) \;.
\label{biaseq}
\ee
The flux power spectrum obtained from the hydro-simulations is used to
determine $b(k)$.  In reality, some mode coupling is expected (Gnedin
\& Hamilton 2002) and the relation between flux and matter power
spectrum will in general not be linear. However, if the true matter
power spectrum is close to the matter power spectrum used to determine
$b(k)$ the approximation has shown to give accurate results.  We refer to
C02, Gnedin \& Hamilton (2002) and Zaldarriaga, Scoccimarro \& Hui
(2003) for more details and for an extensive discussion of the
possible limitation of this method.  

With our grid of models of varying slope and amplitude of the DM power
spectrum, we always have a simulation which comes close to reproducing
the observed flux power spectrum to within $\sim 10 $\% in the range
$0.003 < k$ (s/km) $< 0.03$.  We then use the $b(k)$ determined from
this simulation to infer the `observed' linear matter power spectrum by
dividing the observed flux power spectrum by $b^{2}(k)$.  The
`corrections' of the linear power spectrum with respect to the one
actually simulated are smaller than 10\%.

\begin{figure*}
\center\resizebox{1.0\textwidth}{!}{\includegraphics{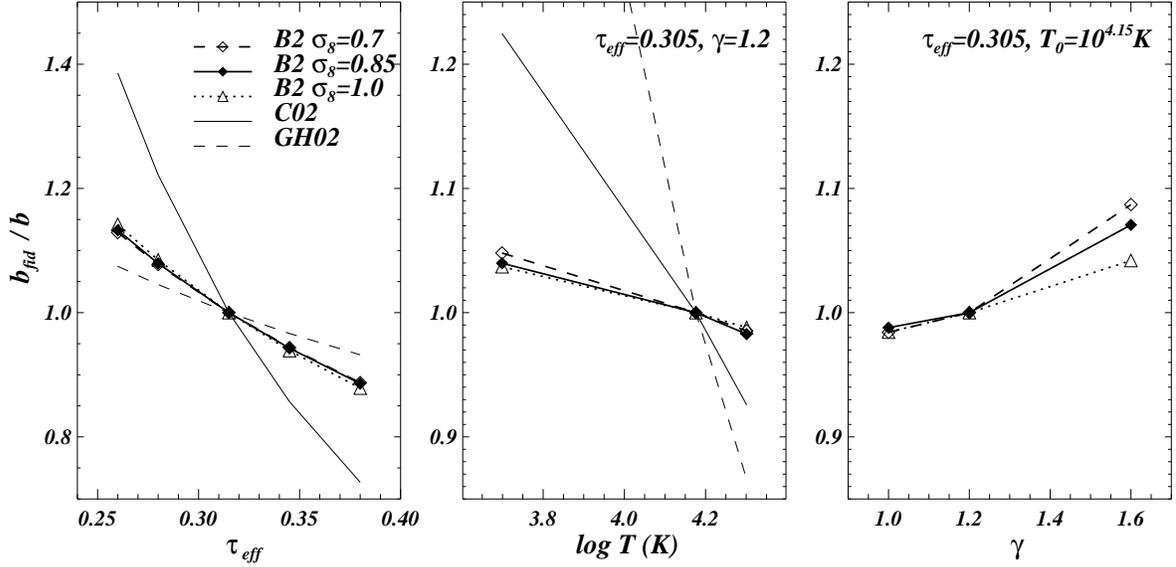}}
\caption{$b_{\rm fid}/b$ for three different simulations with different
values of $\sigma_8$.  {\it Left}: $b_{\rm fid}/b$ as a function of
$\tau_{\rm eff}$ (no scaling of the temperature-density relation has
been adopted here). We also show the scaling found by Croft et
al. (2002) [C02] and Gnedin \& Hamilton (2002) [GH02] as thin
solid and dashed curves, respectively. {\it Middle}: $b_{\rm
fid}/b$ as a function of $T_0$, the temperature at the mean density,
for a fixed $\tau_{\rm eff}=0.305$ and $\gamma=1.2$. {\it Right}:
$b_{\rm fid}/b$ as a function of $\gamma$, the power-law index of the
temperature-density relation, for a fixed $\tau_{\rm eff}=0.305$ and
$T_0=10^{4.15}$ K.}
\label{fig4}
\end{figure*}

\subsection{Statistical errors}
\label{statistical}

As discussed in Section~2.4, the systematic uncertainties on the
observed flux power spectrum are small for 0.003 s/km $< k <$ 0.03
s/km, the range of wavenumber we have used for our analysis. The
errors of the observed flux power spectrum in this range of
wavenumbers should be dominated by statistical errors.  In the
following, we will assume the statistical errors of the inferred
linear matter power spectrum to be the same as the statistical errors
of the observed flux power spectrum. Cosmic variance in the flux power
spectrum of our simulated spectra is a further possible source of
statistical error in our modelling.  However, our analysis is performed
using simulations with a box size of $60$\mpch which corresponds to
6082.3 km/s in velocity space. Our simulation thus probes a volume
which is $\sim 27$ times larger than $(2\pi/k_{\rm min})^3$, where
$k_{\rm min} = 0.003\, {\rm s/km}$ is the smallest wavenumber we use
for our analysis. The cosmic variance error due to the finite size of
our simulation box should thus only moderately affect the smallest
wavenumber we use.  Its contribution to the total error of the rms
fluctuation amplitude will be negligible compared to the other
systematic errors which we describe in the next section.

\subsection{Systematic errors} 
\label{systematics}

\subsubsection{The effective optical depth}

The inferred amplitude of the matter power spectrum depends strongly
on the assumed effective optical depth of the absorption spectrum
(e.g. Croft et. al. 1998; C02; Gnedin \& Hamilton 2002; Seljak,
McDonald \& Makarov 2003).  In Figure \ref{fig4} (right panel), we show
the effect of changing the mean optical depth on the 1D flux power
spectrum.  The dotted curve shows the difference of the 1D flux power
spectrum if $\tau_{\rm eff}$ is changed from 0.305 to 0.349 for the B2
simulation.  Increasing the mean optical depth increases the flux
power spectrum by a factor which is nearly constant for $k< 0.03$
s/km.  

To quantify the effect of the effective optical depth on the inferred
linear matter power spectrum, we consider the ratio $b_{\rm fid}/b$,
where $b_{\rm fid}$ is the average value of the bias function $b(k)$
for a fiducial model in the range considered, and $b$ is the average
value for a model with different parameters ($\tau_{\rm eff}$, $T_0$
and $\gamma$).  In Figure \ref{fig4} (left panel), we show the effect
of changing the assumed effective optical depth on the 3D flux power
spectrum at small $k$, for three different simulations with three
different values of $\sigma_8$ at $z=2.75$.  The dependence on
$\sigma_8$ is weak.  The dependence
of $b_{\rm fid}/b$ on $\tau_{\rm eff}$ is well fitted by \be {b_{\rm
fid} \over b} \sim \left({\tau_{\rm eff}\over 0.305}\right)^ {-0.7}.
\ee This dependence is intermediate between that found by C02 and that
by Gnedin \& Hamilton (2002).

\subsubsection{The temperature-density relation} 
 
The solid, dotted and dashed curves in Figure 3b show the
effect of changing $\gamma$ on the 1D flux power spectrum. Increasing
$\gamma$ leads to a decrease of the flux power, again by a factor
which is nearly constant for $k< 0.03$ s/km.  In the middle and right
panels of Figure \ref{fig4}, we show the dependence of $b_{\rm fid}/b$
on $\gamma$ and $T_{0}$ (averaged over $k$ for $0.003<k/{\rm
(s/km)}<0.03$) where $b_{\rm fid}$ is again the ratio of flux to
matter power spectrum for our fiducial model. 
We fixed  $\tau_{\rm eff}= 0.305$ and  rescaled to $\gamma = 1.2$ (middle
panel) and $T_0=10^{4.15}$ K (right panel), respectively.  Rescaled
'hot' simulations were used  in all cases. 

The dependences of $b_{\rm fid}/b$ on $T_{0}$ and $\gamma$ are fitted
by 
\be {b_{\rm fid} \over b} \sim \left[{1+T_0/(10^{4.15}\,{\rm K}) \over
2}\right]^ {-0.15}, 
\ee and \be {b_{\rm fid} \over b} \sim
\left({1+\gamma/1.3 \over 2}\right)^ {0.3}, \ee respectively.  C02
found a somewhat weaker dependence on the value of $T_0$, and no
dependence on $\gamma$.

\subsubsection{Other systematic errors} 
Other possible systematic errors include fluctuations of the UV
background and potentially modifications of the forest by galactic
winds. Variations of the flux of hydrogen ionising photons are
expected to be large in the intermediate aftermath of
reionisation. However, at the redshifts we are interested in here, they
are expected to be small due to the large mean free path of ionising
photons.  Meiksin \& White (2003) estimate that their effect on the
flux power spectrum should be smaller than 0.5\% at $z=2.75$.  

The effect of galactic winds on the flux power spectrum will depend
strongly on the volume filling factor of galactic outflows and on when
the winds occur (Theuns et al. 2002, Bruscoli et al. 2002).  In the range of wavenumbers
which we used for our analysis, Croft et al. (2003) and Desjaques et
al. (2004) found a negligible effect on the flux power spectrum for
galactic winds consistent with the correlation between flux decrement
and galaxies detected by Adelberger et al. (2003) at $z\sim 3$, while
Weinberg et al. (2003) found a decrease of the flux power at large
scales when considering strong winds models.
Galactic winds appear to affect the flux power spectrum mainly
by their effect on the strong absorption systems which contribute
significantly to the flux power spectrum (Viel et al. 2004a).

\begin{figure*}
\center\resizebox{1.0\textwidth}{!}{\includegraphics{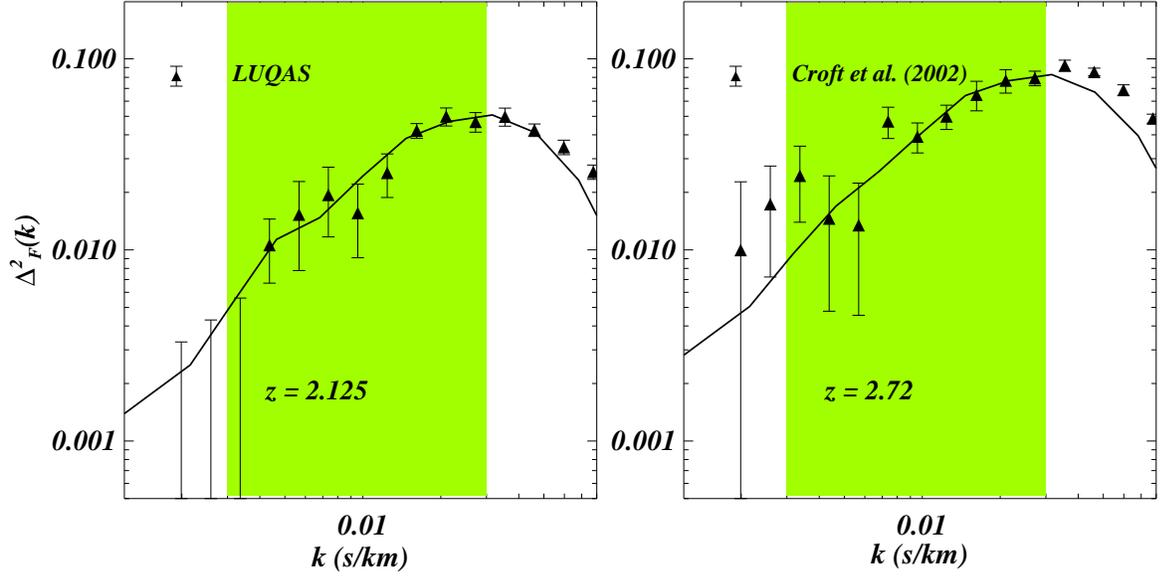}}
\caption{{\it Left}: Flux power spectrum at $z=2.125$. The filled
triangles are for the LUQAS subsample with a mean redshift of
$z=2.125$ (see Table~\ref{tab1}). The continuous curve is the power
spectrum of the simulation that fits the data best.  {\it Right}: Flux
power spectrum at $z=2.72$. The filled triangles are for the fiducial
sample of Croft et al.~(2002). The continuous curve is the power
spectrum of the simulation that fits the data best (with $\tau_{\rm
eff}=0.305$). Shaded regions indicate the range of wavenumbers used in
our analysis.}
\label{fig5}
\end{figure*}

\section{The inferred   matter power spectrum}
\label{linear}
\subsection{The linear power spectrum from LUQAS and Croft et al.}
\begin{figure*}
\center\resizebox{1.0\textwidth}{!}{\includegraphics{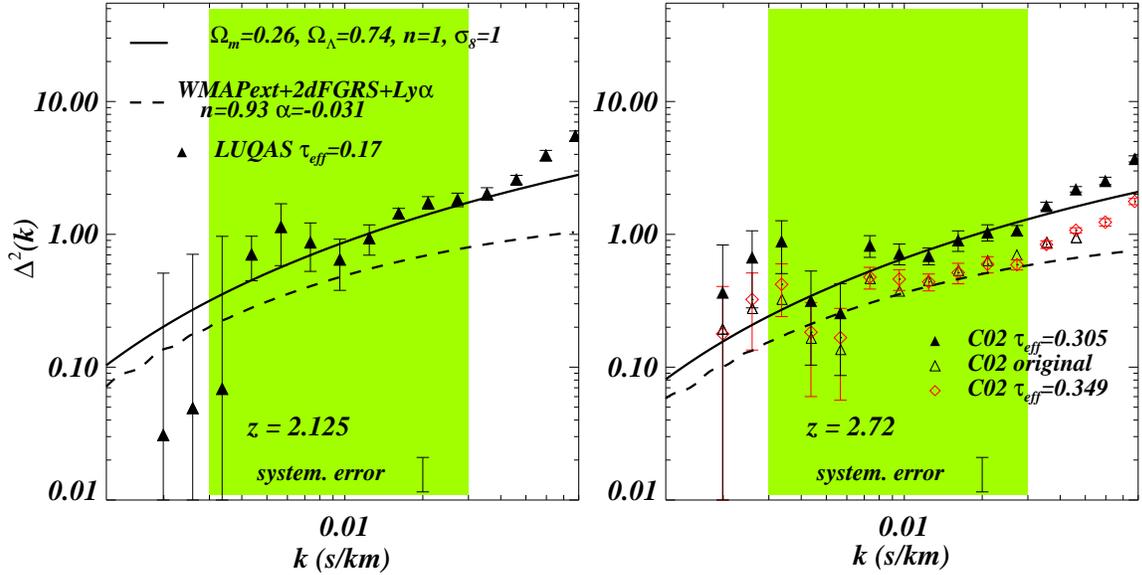}}
\caption{ {\it Left}: Linear power spectrum at $z=2.125$. The linear
dark matter power spectrum inferred from the LUQAS fiducial sample
with the bias parameter derived from a simulation of model C3 which
fits the data best.  The statistical errors are the same as those of
the flux power spectrum in Fig.~\ref{fig5}.  The curve is a
theoretical linear power spectrum with parameters as given on the
plot. The error bar at the bottom of the plot is our estimate of the
systematic uncertainty as discussed in Section~5.2. Shaded regions
indicate the range of wavenumbers used in our analysis.  {\it Right}:
Linear power spectrum at $z=2.72$.  Empty triangles show the result
obtained by C02.  Empty diamonds are the result of our analysis for
the same optical depth as used by Croft et al. ($\tau_{\rm eff} =
0.349$).  Solid triangles are the result of our analysis for our
preferred effective optical depth of $\tau_{\rm eff} = 0.305$.  The
solid and dashed curves are theoretical linear power spectra with
parameters as indicated in the plot. The dashed curve is the ``best
fitting'' model with a running spectral index found by the WMAP team
(Spergel et al. 2003).}
\label{fig6}
\end{figure*}

In this Section, we present the main result of the paper, the linear
dark matter power spectrum we infer for the LUQAS and C02 samples.  As
discussed in Section~\ref{method}, we first have to find the
simulation output for which the 3D flux power spectrum fits the
observed flux power spectrum best. To this end we have used $\chi^2$
minimisation in the range $0.003 < k $ (s/km) $< 0.03$.  For these
wavenumbers the covariance matrix is reasonably close to diagonal,
allowing us to neglect correlations between the data points.  

In Figure \ref{fig5}, we show the flux power spectrum of the  best
fitting simulation  and compare to the observed flux power spectrum 
for the LUQAS sample (rescaled simulation C3 with $\sigma_8 = 1.04$) 
and for the C02 sample (rescaled simulation C3 with $\sigma_8 = 1.035$).  
As discussed in Section~2.4, the assumed effective optical 
depths were $\tau_{\rm eff} = 0.17$ for the LUQAS sample and $\tau_{\rm
eff} = 0.305$ for the
C02 sample.  We then determined the bias function $b(k)$ from the
best-fitting models and use Eqn.~(\ref{biaseq}) to infer the linear
dark matter power spectrum.  
Using the rescaling method of simulations
at different redshifts (see Section~\ref{grid}), we obtained a
sufficiently fine grid in fluctuation amplitude of the matter power
spectrum. There was thus no need to interpolate the flux power spectrum
between simulations of different fluctuation amplitude. 
Table 4 gives the inferred 3D linear dark matter power spectrum 
obtained with $b(k)$ from the best-fitting simulation in Figure 
\ref{fig5} for the LUQAS and Croft et al.  sample, respectively.
The solid triangles in the left and right panels of Figure \ref{fig6}
show  the corresponding $\Delta^2_{\rm mat}=P_{\rm mat}(k)\,k^3/(2\pi^2)$. 
The errors given  are the statistical errors of the flux power spectrum. We will
discuss systematic errors in the next section.  
Our estimate of the DM power spectrum from the C02 sample is $\sim
45\%$ higher than the original result of C02, which is shown as the
open triangles.
The original result of C02 was obtained with a
significantly larger effective optical depth of $\tau_{\rm eff}=0.349$
instead of $\tau_{\rm eff}= 0.305$.  The open
diamonds show our re-analysis of the C02 data with the same effective
optical depth of $\tau_{\rm eff}=0.349$ as C02 have used. There is
good agreement which is remarkable considering the fact that we have
used hydro-dynamical simulations rather than DM simulations and that
the cosmological model is significantly different ($\Omega_{\rm
0m}=0.4$ {\it vs.} $\Omega_{\rm 0m}=0.26$).  Note, however, that this
agreement is somewhat fortuitous given that we have derived a
significantly different scaling of $b(k)$ with $\tau_{\rm eff}$ from
our simulations than C02 do.  For our preferred value of $\tau_{\rm
eff} = 0.305$, C02 would have obtained a 25\% larger amplitude of the
power spectrum than we do.   The solid  curves represent a model 
with $n=1,\sigma_{8}=1$. The dashed curve is the ``best fitting''
model with a running spectral index found by the WMAP team for a
combination of CMB, galaxy survey
and \lya forest data (Spergel et al. 2003). This model falls significantly
below the DM power spectrum which we have inferred from the LUQAS and
the C02 sample.

\begin{table}
\caption{3D linear dark matter power spectrum inferred from the LUQAS 
and  Croft et al. (2002) sample. The Croft et al. sample 
was analysed with $\tau_{\rm eff}=0.305$. The values 
are for  $\gamma=1.6$ and $T_0 = 10^{4.15} $K.}
\label{tab6}
\begin{tabular}{lcc}
\hline
\noalign{\smallskip}
$k$ (s/km) & \multicolumn{2}{c}{$P_{\rm mat}(k)$(km/s)$^{3}$}\\
&&\\
&z=2.125&z=2.72\\
\noalign{\smallskip}
\hline
\noalign{\smallskip}
 0.00199 &    $(7.74 \pm 120)\times 10^7 $  &    $(9.14 \pm 11.67)\times 10^8 $          \\
 0.00259 &    $(5.64 \pm  75.1)\times 10^7 $ &    $(7.67 \pm 4.47)\times 10^8 $         \\
 0.00336 &    $(3.56 \pm  46.7)\times 10^7 $ &    $(4.61 \pm 1.98)\times 10^8 $         \\
 0.00436 &    $(1.68 \pm   0.62)\times 10^8 $ &    $(7.52 \pm 5.06)\times 10^7 $        \\
 0.00567 &    $(1.23 \pm   0.61)\times 10^8 $  &    $(2.77 \pm 1.84)\times 10^7 $       \\
 0.00736 &    $(4.31 \pm   1.71)\times 10^7 $  &    $(4.07 \pm 0.76)\times 10^7 $       \\
 0.00956 &    $(1.46 \pm     0.61)\times 10^7$	&    $(1.62 \pm 0.29)\times 10^7 $      \\
 0.01242 &    $(9.64 \pm     2.48)\times 10^6$ 	&    $(7.09 \pm 1.03)\times 10^6$      \\
 0.01614 &    $(6.77 \pm     0.59)\times 10^6$  &    $(4.24 \pm 0.75)\times 10^6 $      \\
 0.02097 &    $(3.71 \pm     0.40)\times 10^6$  &    $(2.22 \pm 0.31)\times 10^6 $   	  \\
 0.02724 &    $(1.78 \pm     0.21)\times 10^6$   &    $(1.05 \pm 0.91)\times 10^6 $     \\
 0.03538 &    $(9.02 \pm     0.98)\times 10^5$   &    $(7.29 \pm 0.48)\times 10^5 $     \\
 0.04597 &    $(5.25 \pm     0.40)\times 10^5$   &    $(4.44 \pm 0.21)\times 10^5 $     \\
\noalign{\smallskip}
\hline
\end{tabular}
\end{table}

\subsection{The error budget}

In Table~\ref{tab7}, we give a summary of the different sources of
errors for the inferred rms fluctuation amplitude of the matter
density.  The statistical error given in Table~\ref{tab7} is that due
to the statistical errors of the observed flux power spectrum.  The
systematic uncertainties due to $\tau_{\rm eff}$, $\gamma$ and $T_{0}$
are taken from Figure \ref{fig4}.  The estimate of the systematic
uncertainty due to the method is based on Figure~\ref{fig3}. Note 
that due the weak dependence of the inferred amplitude on temerature
which we find the wide adopted possible temperature range nevertheless 
leads to a small error due to the uncertainty in  $T_{0}$.

The uncertainty due to numerical simulations is difficult to estimate,
but we believe that we have demonstrated that we have sufficient
resolution to produce convergent results.  The dot-dashed curve in the
middle panel of Figure~\ref{fig3} shows the effect on the flux power
spectrum of a further increase of the mass resolution by a factor of
eight (this curve can be compared with the similar plots in Figure
2). For the relevant wavenumbers the difference is less than
2\%. However, the discrepancy of the scaling of $b(k)$ with $\tau_{\rm
eff}$ compared to C02 is worrying.  We have thus assigned half of the
difference between their and our result for $\tau_{\rm eff}=0.305$ as
systematic uncertainty due to numerical simulations. This is admittedly
arbitrary and merits further investigation.

Further unknown systematic uncertainties are obviously impossible to
quantify. We have nominally assigned 5\% to take into account a
possible effect due to galactic winds. This could be larger especially
at low redshift for the LUQAS sample, where less is known
observationally about the effect of galactic winds. The sum of the
systematic errors (added in quadrature) for the fluctuation amplitude
is 14.8\% at $z=2.125$ and 14.3\% at $z=2.72$, and is shown as the error
bars at the bottom of Figure~\ref{fig6}.
  
\begin{table}
\caption{Error budget for the determination of the 
rms fluctuation amplitude of the matter density field.}
\label{tab7}
\begin{tabular}{lcl}
\hline
\noalign{\smallskip}
statistical error&\ \ \ \ \ &4\%\\
systematic errors&\ \ \ \ \ &\\
\ \ \ $\tau_{\rm eff}(z=2.125) = 0.17\pm 0.02$& &8\%\\
\ \ \ $\tau_{\rm eff}(z=2.72) = 0.305\pm 0.030$& &7\%\\
\ \ \ $\gamma=1.3 \pm 0.3$&&4\%\\
\ \ \ $T_{0} = 15000\,{\rm K} \pm 10000\,{\rm K} $&&3\%\\
\ \ \ method&&5\%\\
\ \ \ numerical simulations&&8\% (?)\\
\ \ \ further systematic errors&&5\% (?)\\
\noalign{\smallskip}
\hline
\end{tabular}
\end{table}

\subsection{Comparison with previous estimates}

\begin{figure}
\center\resizebox{.5\textwidth}{!}{\includegraphics{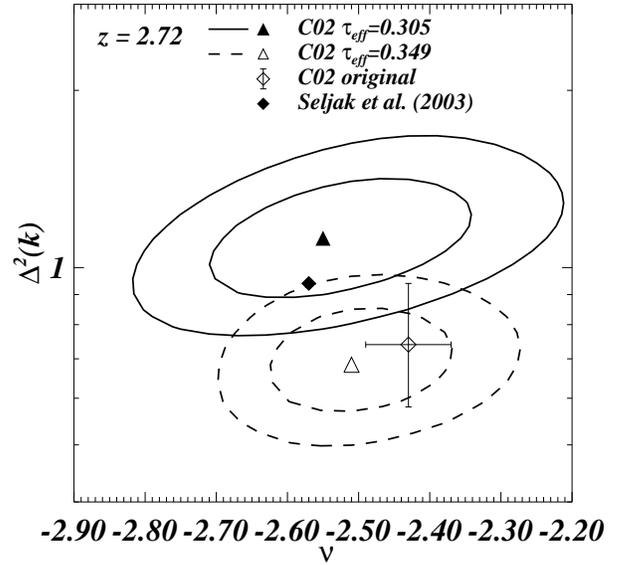}}
\caption{Constraints on the slope and amplitude of the linear dark
matter power spectrum for the fiducial sample of C02 ($z=2.72$).  The
solid contours show the 68.3\% and 95.4\% confidence levels for our
analysis with $\tau_{\rm eff}=0.305$, while the dashed contours are
for $\tau_{\rm eff}=0.349$.  The empty diamond shows the original
result obtained by C02 with $1\sigma$ error bars. The filled diamond
shows the best fitting values obtained by Seljak, McDonald \& Makarov
(2003) (for $\tau_{\rm eff}=0.30$).}
\label{fig7}
\end{figure}

In Figure~\ref{fig7}, we compare the constraints on the amplitude and
slope of the power spectrum with previous estimates.  We fit the
linear dark matter power spectrum with a power-law function $P(k)= P_p
(k/k_p)^{\nu}$ with $k_p=0.03$ s/km, as in C02. We used a diagonal
likelihood for this estimate and convolved the likelihood in the $P_p$
direction with a Gaussian function to take the
systematic errors into account (eqn. 15 of C02).  The contour levels show the 68\% and
95\% confidence levels in the amplitude-slope plane for $\tau_{\rm
eff}=0.305$ (continuous line) and $\tau_{\rm eff}=0.349$ (dashed
line), respectively. 

The empty diamond with error bars is the determination by C02, which
is in good agreement with our determination for the same $\tau_{\rm
eff}$.  As expected, the inferred amplitude increases significantly if
the assumed $\tau_{\rm eff}$ is reduced.  The result of
Seljak, McDonald \& Makarov (2003) for $\tau_{\rm eff}=0.3$ (their
Fig.~1) is also shown as the filled diamond and agrees well with our
determination. Note, however, that the Seljak et al.~result has
been obtained differently. Seljak et al.~fitted a
large grid of flux power spectra obtained from HPM
(Hydro-Particle-Mesh, see Gnedin \& Hui 1998; Meiksin \& White 2001)
simulations with six free parameters to the observed flux power
spectrum of the C02 sample.

\begin{figure*}
\center\resizebox{1.0\textwidth}{!}{\includegraphics{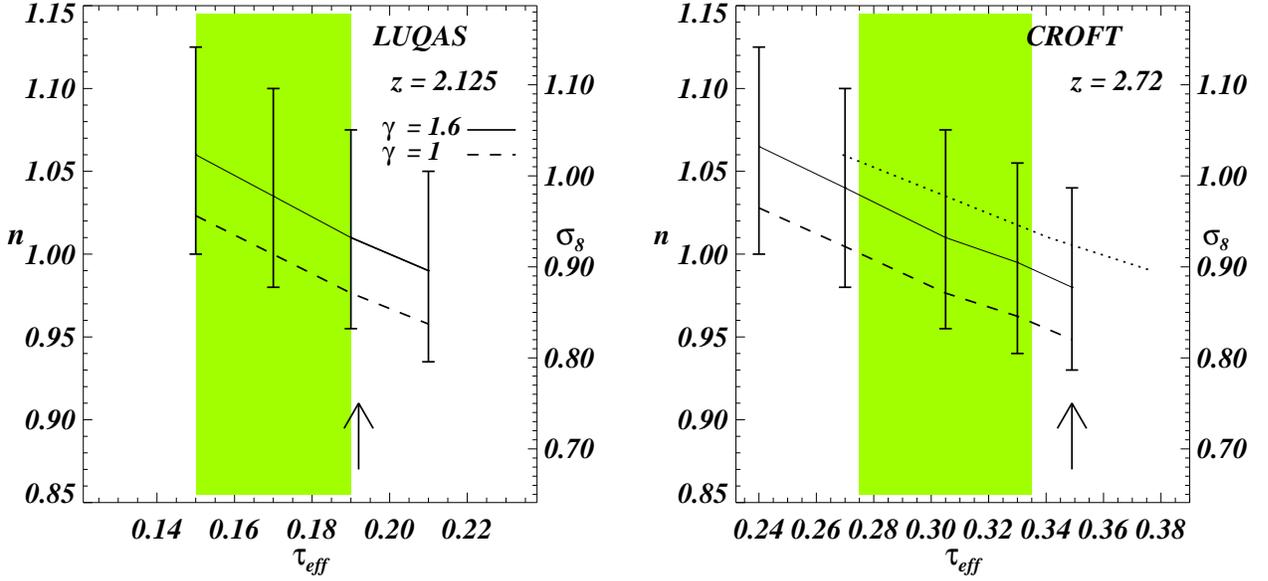}}
\caption{{\it Left}: The values of $\sigma_8$ and $n$ of the best
fitting COBE normalized linear power spectra as a function of the
assumed effective optical depth, for the linear power spectrum
inferred from LUQAS at $z=2.125$.  The solid curve is obtained with
numerical simulations with $\gamma\approx 1.6$ while the dashed curve
is for simulations which were rescaled to have a density temperature
relation with $\gamma = 1.0$. Shaded regions indicate our preferred
values for the effective optical depth as obtained from high-resolution
spectra. The arrow indicates the estimate obtained by Press, Rybicki
\& Schneider (1993) from low-resolution spectra.  {\it Right}: The
same for our re-analysis of the Croft et al.~(2002) data at $z=2.72$.
The dotted line in the right panel is the lower redshift result
scaled by a factor 0.305/0.17 in the $x$-axis.}
\label{fig8}
\end{figure*}

\subsection{Combining CMB and \lya forest data to constrain $n$ and $\sigma_{8}$}
\label{sig8impl}

The measurement of the amplitude of the matter power spectrum on
scales of a few Mpc with the \lya forest is a powerful tool to
constrain the spectral index of primordial density fluctuations when
combined with measurements on large scales from CMB fluctuations (see
Phillips et al. 2001 for a detailed discussion).  The linear matter
power spectrum at $z=0$ can be written as, \be P_{\rm mat} (k) = A
\,k^{n}\, T^2(k), \ee where $T(k)$ is the matter transfer function
which depends on cosmological parameters in the usual way.

To get a first idea how the spectral index inferred by such a combined
analysis depends on $\tau_{\rm eff}$, we here assume a COBE normalized
power spectrum and no contribution by tensor fluctuations. 
We then find the best fitting spectral index $n$ (having fixed
the other cosmological parameters to  our fiducial values). We
use CMBFAST to calculate the theoretical linear power spectra (Seljak
\& Zaldarriaga 1996). Note that normalizing to the WMAP data would
give similar results. 
The thick solid curve in Figure \ref{fig8}
shows the result for a range of $\tau_{\rm eff}$ using numerical
simulations which have $\gamma \approx 1.6$ and $T_{0}\approx
10^{4.15}$ K. The left and right panel are again for the LUQAS and C02
sample, respectively.  The shaded regions indicate the range of
$\tau_{\rm eff}$ preferred by high-resolution absorption spectra as
discussed in Section \ref{effopt}.  For the errors we have added the
statistical errors and the systematic errors in Table~\ref{tab7} in
quadrature. Note that in Figure \ref{fig8} we have omitted the error 
due to $\tau_{\rm eff}$ and $\gamma$ as the dependence on these
parameters is shown explicitely.

Once the spectral shape and the cosmological parameters are fixed, the
measurement of the fluctuation amplitude on \lya forest scales can
also be expressed in terms of $\sigma_{8}$, the rms fluctuation
amplitude of the density for a 8 \mpch sphere. The corresponding
values are shown on the right axis of Figure \ref{fig8}. Note that the
scale corresponding to 8 \mpch is $\sim 0.008\,{\rm s/km}$ at $z=2.72$
(assuming again the cosmological parameters of Section \ref{hydro}).
The flux power spectrum is thus a direct probe of $\sigma_8$ for the
range of $k$-values used in our analysis.

\begin{figure}
\center\resizebox{.5\textwidth}{!}{\includegraphics{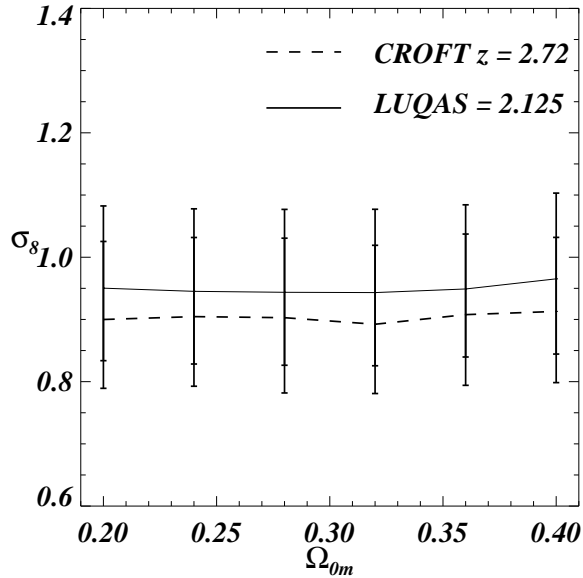}}
\caption{The value of $\sigma_8$ of the best fitting COBE normalized
linear power spectra as a function of the assumed $\Omega_{\rm 0m}$
for the linear power spectrum, inferred from LUQAS at $z=2.125$, and
from the C02 data at $z=2.72$.  The assumed effective optical depths
are  $\tau_{\rm eff} = 0.17$ and $\tau_{\rm eff}=0.305$  and the
exponent of the temperature-density relations was  $\gamma=1.3$.}
\label{fig9}
\end{figure}

The inferred values of $\sigma_{8}$ and $n$ will also depend on the
assumed cosmological parameters (see Phillips et al. 2001 for a
detailed discussion).  Most important is the dependence on
$\Omega_{\rm 0m}$ and $h$.  For a flat cosmological model, the spectral
index scales approximately as $n \propto (\Omega_{\rm 0m}h^2)^{-0.35}$
(Phillips et al. 2001). In
Figure~\ref{fig9}, we show the depencence of the inferred value of
$\sigma_{8}$ on $\Omega_{\rm 0m}$ for a flat universe. This dependence is
essentially negligible.  The dependence on $h$ is also weak.  For the
error estimate we have added all values in Table~\ref{tab7} in
quadrature. Table~\ref{tab9} lists our final estimate of $\sigma_{8}$
and $n$ for the LUQAS and C02 sample, as well as their weighted mean.

\begin{table*}
\caption{Estimated values of $\sigma_{8}$ and $n$ and their
  errors ($\gamma=1.3$ and $\tau_{\rm eff}=0.17$ and $\tau_{\rm
eff}=0.305$  at $z=2.125$ and at $z=2.72$. Note that $n$ scales as $n \propto
  (\Omega_{\rm 0m}h^2/0.135)^{-0.35}$ $^{(\mathrm{a})}$.}
\label{tab9}
\begin{tabular}{cccc}
\hline
\noalign{\smallskip}
&LUQAS& C02 & combined\\
\noalign{\smallskip}
\hline
\noalign{\smallskip}
$\sigma_{8}$&$0.95  \pm 0.04 {(\rm stat.)}\pm 0.13{(\rm
syst.)} $&$0.92  \pm 0.04 {(\rm stat.)}\pm 0.13{(\rm
syst.)} $&$ 0.93  \pm 0.03 {(\rm stat.)}\pm 0.09{(\rm
syst.)} $\\
$n$&$1.02  \pm 0.02 {(\rm stat.)}\pm 0.08{(\rm
syst.)} $&$0.99  \pm 0.02 {(\rm stat.)}\pm 0.08{(\rm
syst.)} $&$1.01  \pm 0.02 {(\rm stat.)}\pm 0.06{(\rm
syst.)} $\\
\noalign{\smallskip}
\hline
\end{tabular}
\begin{list}{}{}
\item[$^{(\mathrm{a})}$] Note that there is an additional uncertainty
in the COBE normalization (Bunn \& White 1997).
\end{list}
\end{table*}

\subsection{Gravitational growth} 

Our subsample drawn from the LUQAS sample was chosen to maximise the contrast in
redshift with respect to the C02 sample, and to further investigate
the redshift evolution of the flux power spectrum. There are two
effects which are responsible for the evolution of the flux power
spectrum, the decrease of $\tau_{\rm eff}$ with decreasing redshift,
and the increase of the fluctuation amplitude of the matter power
spectrum due to gravitational growth. As can be seen by comparing the
right and left panels of Figure~\ref{fig5}, the net effect is a decrease
of the flux power with decreasing redshift which is, however,
significantly smaller than that expected from the decrease of
$\tau_{\rm eff}$.  

In order to assess if the flux power spectra of the LUQAS and C02
samples are consistent with the expected gravitational growth of the
matter power spectrum between the two redshifts, we can compare the
inferred values of $\sigma_{8}$, which should then be the same. To
facilitate such a comparison, the dashed curve in the right panel of
Figure \ref{fig8} is the inferred $\sigma_{8}$ of the LUQAS sample
(shown in the left panel) with all effective optical depths scaled by
the same factor of $0.305/0.17$. The inferred values of $\sigma_{8}$
agree to within the errors, and the evolution of the flux power
spectrum between $z=2.7$ and $z=2.1$ is fully consistent with being
due to the expected gravitational growth and the observed evolution
of $\tau_{\rm eff}$. The same was found by C02 when they compared their
fiducial sample to their low-redshift subsample.

\section{Discussion and Conclusions}
\label{conclu}

We have used the observed flux power spectrum of two large samples of
high-resolution spectra, a sample drawn from LUQAS at a median
redshift of $z=2.125$, and the sample compiled by Croft et al. (2002)
at $z=2.72$,
together with a suite of high-resolution numerical simulations, to
infer the dark matter power spectrum on scales $0.003< k $(s/km)$<0.03$.

We have obtained the following results: 
\begin{enumerate}
\item{With the same assumptions for effective optical depth,
      density-temperature relation, and cosmology, our inferred linear
      matter power spectrum agrees very well with that inferred by
      Croft et al.~(2002).}

\item{We confirm previous results that the inferred rms amplitude of
      density fluctuations depends strongly on the assumed $\tau_{\rm
      eff}$. It increases by 20\% if we assume an optical depth of
      $\tau_{\rm eff}= 0.305$ a value suggested by studies of high-resolution
      absorption spectra. We find, however, a dependence on $\tau_{\rm
      eff}$ which is weaker than that of C02 and stronger than that of
      GH02.}

\item{For values $\tau_{\rm eff}$ suggested by high-resolution
      absorption spectra the linear power spectrum of the 
      best fitting running spectral index model of Spergel et
      al. (2003) falls significantly below the linear power spectrum 
      inferred from both the LUQAS and the C02 sample}.

\item{The decrease of the amplitude of the flux power spectrum between
      $z=2.7$ and $z=2.1$ is consistent with that expected due to the
      decrease of $\tau_{\rm eff}$ and the increase of the amplitude
      of matter power spectrum due to gravitational growth.}
\item{Our estimate of the systematic uncertainty of the rms
      fluctuation amplitude of the density ($\sim 14.5 \%$) is a factor 3.5
      larger than our estimate of the statistical error ($\sim 4 \%$).  The
      systematic uncertainty is dominated by the uncertainty in the
      mean effective optical depth and --- somewhat surprisingly ---
      by the uncertainties between the numerical simulations of different
      authors. Reducing  the overall errors will thus mainly rely 
      on a better understanding of a range of systematic uncertainties.}
\item{By combining the CMB constraint (assuming that there is no contribution from tensor
      fluctuations) on the amplitude of the DM power
      spectrum on large scale with the high-resolution \lya forest
      data we obtain $n= 1.01 \; (\Omega_{\rm 0m}h^2/0.135)^{-0.35}\; \pm 0.02
      {(\rm statistical)}\pm 0.06{( \rm systematic)}$ for the spectral
      index. The corresponding rms fluctuation amplitude is, $\sigma_8
      = 0.93 \pm 0.03 {(\rm statistical)}\pm 0.09{(\rm systematic)}.$}
\end{enumerate}

\section*{Acknowledgments.} 
This work is based on data taken from the ESO archive obtained with
UVES at VLT, Paranal, Chile as part of the LP programme (P.I.: J. Bergeron)
and it is supported by the European Community Research and Training
Network ``The Physics of the Intergalactic Medium''. 
The simulations
were run on the COSMOS (SGI Altix 3700) supercomputer at the
Department of Applied Mathematics and Theoretical Physics in Cambridge
and on the Sun Linux cluster at the Institute of Astronomy in
Cambridge. COSMOS is a UK-CCC facility which is supported by HEFCE and
PPARC.  We are grateful to Tae-Sun Kim for providing us with the UVES sample.
We thank Bob Carswell, George Efstathiou, Tae-Sun Kim,  Massimo Ricotti
and Jochen Weller for useful discussions and PPARC for financial support.


\begin{thebibliography}{}
\bibitem[]{} Adelberger K. L., Steidel C. C., Shapley A. E., Pettini M., 2003, ApJ, 584, 45 
\bibitem[]{} Bernardi M., et al., 2003, AJ, 125, 32  
\bibitem[]{} Bennett C. L., et al., 2003, ApJS, 148, 1
\bibitem[]{} Bruscoli M., et al., 2003, MNRAS, 343, 51
\bibitem[]{} Bunn E. F., White M., 1997, ApJ, 480, 6
\bibitem[]{} Choudhury T.R., Srianand R., Padmanabhan T., 2001, ApJ, 559, 29
\bibitem[]{} Croft R. A. C., 2003, astro-ph/0310890
\bibitem[]{} Croft R. A. C., Weinberg D. H., Katz N., Hernquist L.,
1998, ApJ, 495, 44
\bibitem[]{} Croft R. A. C., Weinberg D. H., Pettini M., Hernquist L.,
Katz N.,  1999b, ApJ, 520, 1 (C99)
\bibitem[]{} Croft R. A. C., Weinberg D. H., Bolte M., Burles S.,
Hernquist L., Katz N., Kirkman D., Tytler D., 2002, ApJ, 581, 20 (C02)
\bibitem[]{} Desjacques V., Nusser A., Haehnelt M. G., Stoehr F., 2004,
MNRAS, in press
\bibitem[]{} Eisenstein D. J., Hu W., 1999, ApJ, 511, 5
\bibitem[]{} Gnedin N. Y., Hui L., 1998, MNRAS, 296, 44 
\bibitem[]{} Gnedin N. Y., Hamilton A. J. S., 2002, MNRAS, 334, 107 
\bibitem[]{} Hui L., 1999, ApJ, 516, 519
\bibitem[]{} Hui L., Gnedin N., 1997, MNRAS, 292, 27
\bibitem[]{} Hui L., Burles S., Seljak U., Rutledge R. E., Magnier E., Tytler D., 2001, ApJ, 552, 15
\bibitem[]{} Kim, T.-S., Carswell, R. F., Cristiani, S., 
D'Odorico, S., Giallongo, E. 2002, MNRAS, 335, 555
\bibitem[]{} Kim, T.-S., Viel M., Haehnelt M.G., Carswell R.F.,
Cristiani S.,  2004, MNRAS,  347, 355 (K04)
\bibitem[]{} McDonald P., Miralda-Escud\'e J., Rauch M., Sargent W.L.,
Barlow T.A., Cen R., Ostriker J.P., 2000, ApJ, 543, 1 (M00)
\bibitem[]{} McDonald P., 2003, ApJ, 585, 34
\bibitem[]{} Meiksin A., White M., 2001, MNRAS, 324, 141
\bibitem[]{} Meiksin A., White M., 2003, astro-ph/0307289
\bibitem[]{} Press W. H., Rybicki G. B., Schneider D. P., 1993, ApJ,
  414, 64
\bibitem[]{} Phillips J., Weinberg D. H, Croft R.A.C., Hernquist L.,
Katz N., Pettini M., 2001, 560, 15
\bibitem[]{} Rauch M., 1998, ARA\&A, 36, 267
\bibitem[]{} Ricotti M., Gnedin N., Shull M., 2000, ApJ, 534, 41
\bibitem[]{} Schaye J., Theuns T., Rauch M., Efstathiou G., Sargent
W. L. W., 2000, MNRAS, 318, 817
\bibitem[]{} Schaye J., Aguirre A., Kim T.-S., Theuns T., Rauch M.,
Sargent W. L. W., 2003, ApJ, 596, 768
\bibitem[]{} Seljak U., Zaldarriaga M., 1996, ApJ, 469, 437
\bibitem[]{} Seljak U., McDonald P., Makarov A., 2003, astro-ph/0302571
\bibitem[]{} Spergel D. N. et al. 2003, ApJS, 148, 175
\bibitem[]{} Springel V., Yoshida N., White S.~D.~M., 2001, NewA, 6, 79
\bibitem[]{} Springel V., Hernquist L., 2002, MNRAS, 333, 649
\bibitem[]{} Springel V., Hernquist L., 2003, MNRAS, 339, 289
\bibitem[]{} Theuns T., et al., 1998, MNRAS, 301, 478
\bibitem[]{} Theuns T., Viel M., Kay S., Schaye J., Carswell B., Tzanavaris P., 2002, ApJ, 578, L5
\bibitem[]{} Tytler D., et al., 2004, astro-ph/0403688
\bibitem[]{} Verde L., et al. 2003, ApJS, 148, 195
\bibitem[]{} Viel M., Matarrese S., Theuns T., Munshi D., Wang Y.,
2003, MNRAS, 340, L47
\bibitem[]{} Viel M., Haehnelt M. G., Carswell R. F., Kim T.-S., 2004a,
MNRAS, 349, L33
\bibitem[]{} Viel M., Matarrese S., Heavens A., Haehnelt M. G., Kim
T.-S., Springel V., Hernquist L., 2004b, MNRAS, 347, L26
\bibitem[]{} Weinberg D., 1999, in: Evolution of large scale structure
: from recombination to Garching, eds. A. J. Banday, R. K. Sheth,
L. N.  da Costa., p.346
\bibitem[]{} Weinberg D.H., Dave' R., Katz N., Kollmeier J.A., 2003, AIP
conf. Proc. 666, 157-169
\bibitem[]{} Zaldarriaga M., Hui L., Tegmark M., 2001, ApJ, 557, 519
\bibitem[]{} Zaldarriaga M., Scoccimarro R., Hui L., 2003, ApJ, 590, 1
\bibitem[]{} Zuo L., Bond J. R., 1994, ApJ, 423, 73 
~\end{thebibliography}
\end{document}